\documentclass[journal=jacsat]{achemso}
\usepackage[version=3]{mhchem}
\usepackage{hyperref}
\usepackage{etoc} 
\usepackage{color}
\usepackage{units}
\usepackage{float}
\usepackage{ulem}

\definecolor{comment1}{rgb}{0,0,1}

\author{Aaron R. Finney}
\email{a.finney@ucl.ac.uk}
\author{Matteo Salvalaglio}
\email{m.salvalaglio@ucl.ac.uk}
\affiliation[UCL]
{Thomas Young Centre and Department of Chemical Engineering, University College London, London WC1E 7JE, United Kingdom}

\title[]{Multiple Pathways in NaCl Homogeneous Crystal Nucleation}

\begin{document}
\etocdepthtag.toc{mtchapter}
\etocsettagdepth{mtchapter}{subsection}
\etocsettagdepth{mtappendix}{none}

\singlespacing
\begin{abstract}
\noindent NaCl crystal nucleation from metastable solutions has long been considered to occur according to a single-step mechanism where the growth in the size and crystalline order of the emerging nuclei is simultaneous. 
Recent experimental observations suggest that significant ion-ion correlations occur in solution and that NaCl crystals can emerge from disordered intermediates which is seemingly at odds with this established view.
Here, we performed biased and unbiased molecular dynamics simulations to analyse and characterise the pathways to crystalline phases from solutions far into the metastable region.
We find that large liquid-like NaCl clusters emerge as the solution concentration is increased and a wide distribution of crystallisation pathways are observed with two-step nucleation pathways---where crystalline order emerges in dense liquid NaCl regions---being more dominant than one-step pathways to phase separation far into the metastable region.
Analyses of cluster size populations and the ion pair association constant show that these clusters are transient, unlike the thermodynamically stable prenucleation cluster solute species that were suggested in other mineralising systems.
A Markov State Model was developed to analyse the mechanisms and timescales for nucleation from unbiased molecular dynamics trajectories in a reaction coordinate space characterising the dense regions in clusters and crystalline order.
This allowed calculation of the committor probabilities for the system to relax to the solution or crystal states and to estimate the rate of nucleation, which shows excellent agreement with literature values. 
From a fundamental nucleation perspective, our work highlights the need to extend the attribute `critical' to an ensemble of clusters which can display a broad range of structures and include sizeable disordered domains depending upon the reaction conditions. 
Moreover, our recent simulation studies demonstrated that carbon surfaces catalyse the formation of liquid-like NaCl networks which, combined with the observations here, suggests that alternative pathways beyond the single-step mechanism can be exploited to control the crystallisation of NaCl.

\end{abstract}

\section{Introduction}
Many pathways to crystals have been observed in a range of systems that \textit{seemingly} deviate from the single-step crystal nucleation mechanism from a parent solution phase that is predicted by classical nucleation theory (CNT). \cite{de_yoreo_crystallization_2015,zhang_perspective_2020}
It was shown some time ago in simulations of spherical particles with short-ranged interactions that, under appropriate conditions, liquid-like intermediates precede the formation of crystals and the rate for crystal nucleation increases. \cite{wolde_enhancement_1997}
Experiments were later presented that support these observations in systems containing proteins\cite{galkin_control_2000,galkin_liquid-liquid_2002} and small organic molecules. \cite{garetz_polarization_2002,bonnett_solution_2003}
The nucleation of crystals from supersaturated solutions was therefore described as `two-step' \cite{vekilov_two-step_2010} in light of findings such as these. 

Evidence was presented for hydrous mineral dense liquid phases which precede the precipitation of amorphous solid and crystalline phases at room temperature \cite{faatz_amorphous_2004,wolf_early_2008,wang_situ_2013,bewernitz_metastable_2013,smeets_classical_2017,avaro_stable_2020}, demonstrating how multi-step crystallisation pathways are a common feature in a range of crystallising systems. In principle, this observation does not negate CNT as an adequate framework to predict the rates for phase separation in mineralising solutions \cite{hu_thermodynamics_2012,carino_thermodynamic-kinetic_2017}.
Indeed, it should be noted at the outset that a two-step pathway can be described using thermodynamic concepts consistent with CNT \cite{kashchiev_classical_2020} adopting a core-shell composite cluster model \cite{iwamatsu_free-energy_2011,iwamatsu_nucleation_2012}. 
Furthermore, observation of amorphous intermediates alone does not confirm their direct role in crystallisation \textit{i.e.}, crystal nuclei could form in a single-step via stochastic density fluctuations of the dissociated ions in the surrounding lean solution in quasi-equilibrium with amorphous states.
An alternative to the pathway described by CNT posits that liquid-like microscopic associates are \textit{thermodynamically stable} species with respect to dissociated ions and are present even in undersaturated solutions; these were described as `prenucleation clusters' on the basis that they purportedly represent the solute species directly involved in nucleation. \cite{gebauer_stable_2008,demichelis_stable_2011,sebastiani_water_2017,scheck_molecular_2016}
The prenucleation cluster pathway was described as a paradigmatic example of `nonclassical' nucleation because here, aggregation of the stable clusters is the limiting step to phase separation. \cite{gebauer_stable_2008,gebauer_pre-nucleation_2014,avaro_stable_2020}

In NaCl(aq), which is arguably the simplest mineralising solution, computer simulations have been and continue to be instrumental for our understanding of crystal nucleation. \cite{jiang_forward_2018,jiang_nucleation_2019,zimmermann_nucleation_2015,zimmermann_nacl_2018,alejandre_ions_2007,giberti_transient_2013,lanaro_birth_2016,chakraborty_how_2013,zahn_atomistic_2004,pulido_lamas_homogeneous_2021,karmakar_molecular_2019} 
Molecular dynamics simulations using the classical force field of Joung and Cheatham \cite{joung_determination_2008} identified that the nominal equilibrium molality for a saturated NaCl(aq) solution under standard conditions was 3.7~$m$ \cite{moucka_molecular_2013,moucka_chemical_2015,mester_mean_2015,mester_temperature-dependent_2015,benavides_consensus_2016,espinosa_calculation_2016} (where $m$ refers to mol/kg units) and the limit of solution stability was identified at 15~$m$.\cite{jiang_nucleation_2019} 
This provides a relatively wide range of concentrations where the solution is metastable and crystal nucleation is favoured but does not occur spontaneously. 

In metastable NaCl(aq) solutions at $T=298$~K, crystal nucleation rates were determined using seeded molecular dynamics simulations and classical nucleation theory (CNT). \cite{zimmermann_nucleation_2015,zimmermann_nacl_2018,pulido_lamas_homogeneous_2021}
The rates calculated by Pulido Lamas \textit{et al.} were found to match well with those calculated from forward flux sampling (FFS), with both sets of computed rates deviating from the experimentally determined values.\cite{jiang_forward_2018,pulido_lamas_homogeneous_2021}
Zimmerman \textit{et al.} showed that the rates from seeded MD simulations can be aligned to the experimental data by adopting suitable criteria for the identification of ions belonging to the crystal phase. \cite{zimmermann_nacl_2018}
These results thus indicate that CNT provides a consistent model for determining the nucleation kinetics in metastable NaCl(aq) solutions.
At the limit of solution stability, however, simulations identified that, before crystal nucleation, NaCl clusters with minimal crystalline order display large fluctuations in size. \cite{jiang_nucleation_2019}
This informed a two-step description for nucleation of NaCl at and beyond the spinodal, which marks the transition for a change in the type of nucleation mechanism. \cite{jiang_nucleation_2019,pulido_lamas_homogeneous_2021}
The recent experimental observation and characterisation of hydrated dense liquid NaCl domains in supersaturated solutions puts this schematic picture into question. 
In Ref. \citenum{hwang_hydration_2021}, Hwang \textit{et al.} measured long-lived liquid-like nanometre-sized ion domains in levitated droplets by \textit{in situ} X-ray and Raman spectroscopy, and further supported their findings with results from molecular dynamics (MD) simulations.
The formation of these domains at high supersaturations was described as a possible first step in a two-step nucleation process. 

Direct observation of amorphous NaCl intermediates to crystal nucleation was obtained in studies of NaCl(aq) confined to aminated carbon nanotubes performed using transmission electron microscopy after removal of the solvent from the samples. \cite{nakamuro_capturing_2021}
Before the nucleation of a nanocrystal, disordered NaCl clusters emerge and dissipate multiple times.
The nucleation step involved the formation of a (NaCl)$_{50}$ nanocrystal from a semi-ordered precursor.
While this process appears compatible with a two-step mechanism, the drying of the samples \textit{in vacuo}, presence of the (vibrating) carbon nanotube wall and confinement means that the mechanism exposed here is not directly applicable to the case of crystal nucleation in metastable bulk solutions.
Amorphous NaCl nanoparticles have been stabilised over long times by spray-drying; \cite{amstad_production_2015} the mechanism of nanoparticle formation here is not clear and, given the method of preparation, kinetically arrested amorphous structures could well result from spinodal decomposition. 

Here, we employed swarms of MD simulations, seeded MD simulations and swarms of well-tempered metadynamics \cite{barducci_well-tempered_2008} simulations to study the pathways associated with crystal nucleation in highly supersaturated, metastable NaCl(aq) solutions at room temperature and pressure. We analyse and discuss speciation in solution and the role of liquid-like clusters in order to characterise nucleation mechanisms in the context of those described above and identify the signature for multi-step nucleation in this regime. 

\section{Computational Methods}
\label{sec:methods}

\paragraph{MD Simulations of Solutions}
A series of MD simulations were performed to investigate the structure and dynamics of ions in bulk NaCl(aq) solutions. 
The molality of these solutions was 2, 3, 4, 5.1, 6.1, 7.1, 8.1, 9, 10, 11.1, 12.5, 13.7, 15 and 16~mol~kg$^{-1}$ which provided supersaturations, $S=0.5 - 4.3$, defined as $S=b(\mathrm{NaCl})/b^{eq}(\mathrm{NaCl})$ and where $b^{eq}(\mathrm{NaCl}) = 3.7$~mol~kg$^{-1}$.  \cite{espinosa_calculation_2016}
At the lowest concentrations (2--5~mol~kg$^{-1}$) ions and 4,000 water molecules were randomly placed into a cubic simulation cell; the number of NaCl ion pairs increased from 145 to 370 over this molality range.
In the remaining simulations, the number of ions was fixed at 370 NaCl and the number of water molecules decreased from 4,000 to 1,280 to achieve the desired molality.
Ions and water molecules were assigned random initial positions (ensuring no overlap between atoms) in a cubic simulation cell. 
A short 0.5~ns simulation was then performed in the ($N,p,T$) ensemble to relax the simulation cell volume and system potential energy, before a minimum of 10 ns of production runs were performed at each $S$.
In order to account for slower relaxation times at higher concentrations the simulation time was increased to 20~ns beyond 9~mol~kg$^{-1}$ and 50~ns beyond 11.1~mol~kg$^{-1}$. In this set of simulations, 13.7~mol~kg$^{-1}$ represents a special case as it is at this molality where nucleation pathways were evaluated and, hence, the MD simulation time here was substantially extended, reaching 300~ns.

\paragraph{Simulations at $\mathbf{S=3.7}$}
Two approaches were taken to prepare input configurations for simulations at $b($NaCl$)=13.7~m$ ($S=3.7$) initiated from either a homogeneous solution phase (solution) or crystalline seeds (seeded) in solution.
In the first case, cubic simulation cells were randomly populated with ions and water molecules as described above; ten configurations in total were prepared in this manner.
For seeded simulations, crystalline seeds were extracted from a rock salt supercell.
Pseudo-spherical crystals with diameter, $d=0.6-2.4$~nm (in 0.2~nm increments) and containing $N=7$ to 341 ions were cut from the bulk crystal where no constraint was placed on the charge of the resulting nanocrystals.
The smallest seed was comprised of an octahedral [NaCl$_6]^+$ complex and, hence, carried a high charge density.
The absolute excess charge density ($\rho^{ex}$) in seeds decayed rapidly as the number of ions in the seed ($N$) increased, and this can be reasonably approximated by $\rho^{ex}=1.4N^{-0.6}e$~atom$^{-1}$.
The crystalline seeds were subsequently immersed in a solution such that $b($NaCl$)=13.7~m$ according to the total number of ions the simulation box.
All systems were relaxed during 0.5~ns ($N,p,T$) MD simulations in which the ions in crystalline seeds were fixed to their initial positions. This was followed by 300~ns production runs, where all ions were mobile except for one cation at the centre of the crystal which was fixed to the centre of the simulation cell.
These initial configurations (both solutions and seeds) were also used to perform well-tempered metadynamics simulations, as described in the following section.

\paragraph{Simulation Details}
In line with other recent simulation studies of NaCl nucleation, \cite{zimmermann_nucleation_2015,zimmermann_nacl_2018,jiang_forward_2018,jiang_nucleation_2019} ions in solution were modelled using the Joung and Cheatham force field \cite{joung_determination_2008} which adopts the SPC/E model \cite{berendsen_missing_1987} for water.
The geometry of water molecules was constrained using the LINCS algorithm. \cite{hess_lincs:_1997}
MD simulations were performed using GROMACS 2018.6 \cite{hess_gromacs_2008} and the leapfrog time integration algorithm with a 1~fs timestep. 
The temperature and pressure of the system were held constant at 298~K and 1~bar, respectively, within statistical fluctuations using the Bussi-Donadio-Parrinello thermostat \cite{bussi_canonical_2007} and the barostat of Berendsen \textit{et al.} \cite{berendsen_molecular_1984} 
Particle Mesh Ewald summation \cite{essmann_smooth_1995} was adopted to compute the energies and forces on atoms arising from electrostatic interactions.
As three-dimensional periodic boundary conditions were used throughout, the real-space contributions to the Ewald sum were computed for atoms within 0.9~nm, and Lennard-Jones interactions were truncated at 0.9~nm with a dispersion correction added to the energies of short-range intermolecular interactions. 

\paragraph{Collective Variables}
We analysed pathways from solutions to crystals on a two-dimensional reaction coordinate space characterised by the collective variables (CVs) $n$ and $n(q6)$ which quantify the total size of high ion density regions and crystalline regions in NaCl clusters, respectively.
These calculations were performed using PLUMED (version 2.5).
\cite{tribello_plumed_2014}
Here, we determined the coordination between ions according to a smoothly varying geometric criterion:
\begin{equation}
    f=\frac{1-(\frac{r_{ij}}{r_0})^p}{1-(\frac{r_{ij}}{r_0})^{q}}
    \label{eq:switch}
\end{equation}
where $p=6$, $q=12$, $r_{ij}$ is the distance between two ions $i$ and $j$ and $r_0=0.38$~nm. 
The total coordination number for $i$ was then calculated by shifting and stretching $f_i$, such that the value of coordination goes smoothly to zero at $r_{max}=1$~nm:
\begin{equation}
    c_i=\frac{1}{N_j}\sum_{j=1}^{N_j}\frac{f(r_{ij})-f(r_{max})}{1-f(r_{max})}
    \label{eq:cn}
\end{equation}
where the sum runs over all $j$ ions with opposite charge to $i$.
A variable $s$ is then defined as
\begin{equation}
    s_i = 
    \begin{cases}
    1,  & \text{if } c_i\geq 5\\
    0,  & \text{otherwise}
    \end{cases}
    \label{eq:gt5}
\end{equation}
where a smooth truncation, with similar functional to Equation \ref{eq:switch}, is applied to the continually varying $c_i$ to determine the value of $s_i$. 
The sum over $s$ provides a measure of the number of ions which are significantly dehydrated, due to coordination to at least five other ions of opposite charge: 
\begin{equation}
    n=\left(\sum_{i=1}^{N_i}s_i+\sum_{j=1}^{N_j}s_j \right)^{\nicefrac{1}{3}}
    \label{eq:cvdef}
\end{equation}
We take the cubic root over the sum in Equation \ref{eq:cvdef} to obtain a CV which scales approximately linearly with the radii of clusters emerging in solution---assuming that clusters have a relatively uniform ionic density throughout their volume---and to expand the reaction coordinate space in the region where the smallest clusters are represented.

The spatial distribution of $s$, $\xi(s)$, was calculated in order to identify ions as a function of distance, $r$, from the centre of the cubic simulation cell:
\begin{equation}
    s^{sph} = \xi(s) \left( 1 - \mathrm{tanh} \left( \frac{r}{r_0} \right) \right)
    \label{eq:sphere}
\end{equation}
where $r_0$ is 1.5~nm.
$n^{sph}$ was calculated according to this equation during biased sampling simulations (see below) to limit the number of nucleation events that occur simultaneously in the simulations.

Together with CVs $n$ and $n^{sph}$, which inform about regions of high local ion density, we also determined the degree of crystallinity in these regions.
To this aim, we calculated a local average of the sixth order Steinhardt bond-orientational order parameter \cite{steinhardt_bond-orientational_1983} defined according to,
\begin{equation}
    S_i = \frac{\sum_j f(r_{ij}) \sum_{m=-6}^6 q_{6m}^*(i)q_{6m}(j)}{\sum_j f(r_{ij})}
    \label{eq:lq6}
\end{equation}
where $f$ is provided in Equation \ref{eq:switch} and, as for $s_i$, $i$ and $j$ represent ions with opposite sign of charge. 
The components of the complex $q6$ vector are calculated using the spherical harmonics $Y_{6m}$:
\begin{equation}
    q6(i) = \frac{\sum_j f(r_{ij}) Y_{6m}(\mathbf{r}_{ij})}{\sum_j f(r_{ij})}
    \label{eq:q6}
\end{equation}
A Gaussian filter was then applied to identify ions with local $q_6$ symmetry that matched the rock salt structure:
\begin{equation}
    \sigma_i = 
    \begin{cases}
    1,  & \text{if } S_i\geq 0.55\\
    0,  & \text{otherwise}
    \end{cases}
    \label{eq:gtq6}
\end{equation}
The cut-off of 0.55 was chosen in this function as it correctly identifies ions at the surface of a bulk NaCl crystal as belonging to a crystalline configuration. 
Finally, Equations \ref{eq:cvdef} and \ref{eq:sphere} are applied to calculate $n(q6)$ and $n^{sph}(q6)$.

\paragraph{Well-Tempered Metadynamics}
Well-tempered metadynamics
\cite{barducci_well-tempered_2008} was adopted in this work to enhance the sampling of local fluctuations of the ion density by biasing $n^{sph}$ on the fly during MD simulations using PLUMED. \cite{tribello_plumed_2014}
During the simulations, an artificial bias potential, $V$, was evolved over time by the iterative deposition of Gaussian contributions to $V$ every 1~ps with $\sigma_n=0.01$, initial height, $w_n=2.5$~kJ~mol$^{-1}$, and bias factor, $\gamma=200$. The bias in $n$ was stored on a grid with limits $-1$ to 10 (bin width = 0.002). 
Metadynamics simulations were performed for around 500~ns initiated from twenty independent configurations obtained from either a solution or crystal seed in solution at $S=3.7$, providing a total of $\sim 10$~\textmu{s} of simulation time.

Relative free energies between states can be calculated according to:
\begin{equation}
   F = -k_{\mathrm{B}}T \ln{(\mathbf{p})}
\end{equation}
where $k_{\mathrm{B}}$ is Boltzmann's constant, $T$ is temperature and $\mathbf{p}$ is the probability density of states in the reaction coordinate space.
Metadynamics enhances the sampling of $n^{sph}$, but translation of the NaCl particle outside of the biased sphere can lead to inaccurate estimates for the probability of $n^{sph}$ states.
It was therefore necessary to calculate the free energies between states in a space characterised by $n$ and $n(q6)$.
To do this, we first calculate the biased probability density ($\mathbf{p}^b(n^{sph},n,n(q6)$) by post-processing the simulation trajectories. 
We also calculated $\mathbf{p}(n^{sph},n,n(q6))$ by reweighting the biased probability density according to the weights, $\exp{\left( \frac{V(n^{sph})}{k_{\mathrm{B}}T} \right)}$.
Metadynamics simulations that failed to sample both the solution and crystalline states and transition states between them were neglected, and the remaining 14 data sets were aligned such that the average $\mathbf{p}(n^{sph},n,n(q6))$ for the solution phase was equal within noise.
The relative free energy was then calculated according to the weighted average:
\begin{equation}
   F = \frac{\sum_i \mathbf{p}^b(-k_{\mathrm{B}}T \ln{(\mathbf{p})})}{\sum_i \mathbf{p}^b}
\end{equation}
where the sum runs over all $i$ simulations that met the sampling criteria described above.
$F(n,n(q6))$ was established by taking a thermodynamic according to,
\begin{equation}
    F(n,n(q6)) = -k_{\mathrm{B}}T \ln{} \int_{n^{sph}} \exp \left( \frac{-F(n^{sph},n,n(q6))}{k_{\mathrm{B}}T} \right) dn^{sph}
\end{equation}

\paragraph{Umbrella Sampling}
To investigate ion association, we performed Umbrella Sampling simulations in solutions approaching the limit of infinite dilution.
Here, one Na$^+$, one Cl$^-$ and 4,000 water molecules were inserted into a cubic simulation cell which was relaxed in ($N,p,T$) simulations for 0.5~ns. A total of 20 ($N,p,T$) simulations (windows) followed, where the distance between ions was restrained at $r_{\mathrm{Na-Cl}}^0=0.25$---1~nm in 0.025~nm increments by the introduction of a time-independent harmonic bias potential:
\begin{equation}
    V_{\mathrm{US}} = \frac{k}{2}(r_{\mathrm{Na-Cl}} - r_{\mathrm{Na-Cl}}^0)^2
\end{equation}
where $k$ was 900~kJ~mol$^{-1}$ for $r_{\mathrm{Na-Cl}}<0.4$~nm and 500~kJ~mol$^{-1}$, otherwise.
Simulations were performed for 5.5~ns, with the initial 0.5~ns discarded in any subsequent analyses.
With this setup, a significant overlap of the probability densities in the reaction coordinate between adjacent windows was obtained, enabling a well converged calculation of the potential of mean force, obtained with the weighted histogram analysis method. \cite{grossfield_wham_nodate}


\paragraph{Ion Clusters and Diffusion}
A depth first search algorithm \cite{tribello_analyzing_2017} was used to determine the cluster size distribution according to a continuous but sharp definition of ion coordination in the first sphere using Equations \ref{eq:cn} and \ref{eq:switch} where $r_0=0.355$~nm, $r_{max}=0.36$~nm, $p=6$ and $q=12$.
In addition, self-diffusion coefficients, $D$, for ions and water were calculated from the mean squared displacement of ions according to, $D=\lim_{t\rightarrow{\infty}}d<(\mathbf{r}(t)-\mathbf{r}(0))^2>/6dt$, where $\mathbf{r}(t)$ are the atom positions at time $t$. 
A correction was applied to account for the finite size of the simulation cell\cite{yeh_system-size_2004} and here the shear viscosity of the adopted water model was taken from Ref. \citenum{gonzalez_shear_2010}.

\paragraph{Markov State Model Construction}
All unbiased simulations, totalling more than 20 \textmu{s}, were used to inform 
Markov State Models (MSMs) in the reaction coordinate space defined by CVs $n$ and $n(q6)$, as well as by their localised counterparts $n^{sph}$ and $n^{sph}(q6)$.
MSM construction and analyses were implemented using pyemma 2.5.6 \cite{scherer2015pyemma}.  
The two-dimensional reaction coordinate spaces were partitioned into discrete sets of states via regular space clustering\cite{prinz2011markov,hartigan1975clustering}, leading to sets of approximately 100 states in both ($n$,$n(q6)$) and ($n^{sph}$,$n^{sph}(q6)$) which are fully connected to one another via intermediate states. 
MSMs constructed from these sets showed a convergent behavior for the slowest implied timescale for lag-times $\tau>50 $~ps (see Figures \ref{fig:msm-sph-si} and \ref{fig:msm-box} in SI)\cite{bowman2013introduction}. 
Bayesian MSMs were therefore employed to estimate the average and 95\% confidence interval of the slowest implied timescale using $\tau =100$~ps, and to estimate the equilibrium probability distribution.\cite{trendelkamp2015estimation}
Using discrete transition path theory (TPT)\cite{metzner2009transition,noe2009constructing}, the MSMs were used to compute the committor probability, as well as the state-to-state probability flux and the mean first passage time associated with the nucleation process. 
This analysis, further described in the Results and Discussion section, provides quantitative information on the pathways connecting the liquid and crystal states. 

\section{Results and Discussion}

\subsection{Solution Speciation}
\begin{figure}[H]
    \centering
    \includegraphics[width=0.75\linewidth]{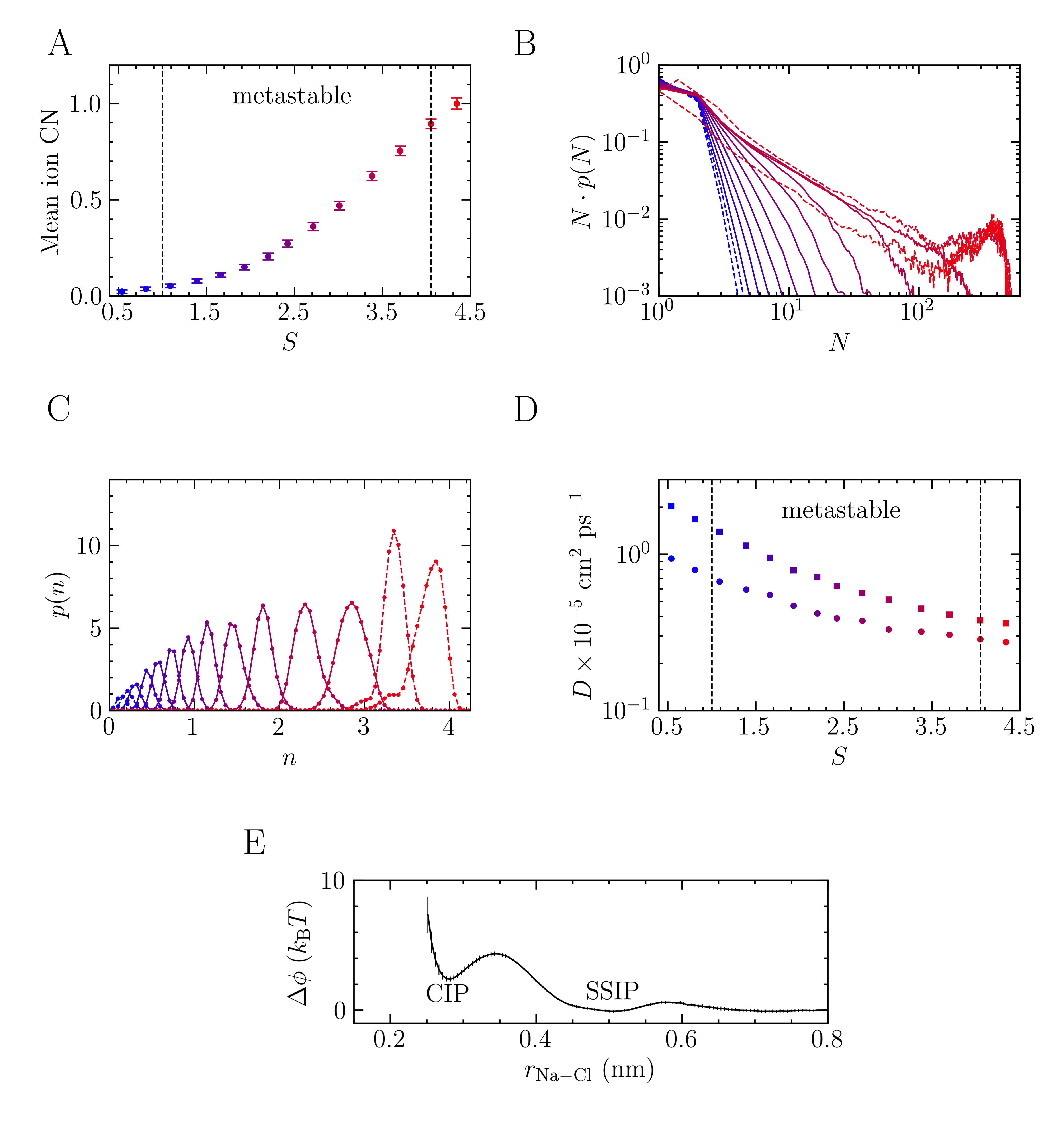}
    \caption{Ion structure and dynamics in solution over a range of supersaturation ratios, $S$. A provides the average ion coordination number in the first coordination sphere. B shows the number weighted cluster size probability distribution ($p(N) \cdot N$) where $N$ is the total number of ions in a cluster defined according to coordination in the first ionic sphere. C provides the $n$ probability densities. D shows the cation (circles) and water (squares) self-diffusion coefficients; the uncertainties in the data are of the size of the data points. The blue $\rightarrow{}$red colour scale in all panels indicates increasing supersaturation from $S=0.5\rightarrow{4.3}$. The dashed lines in A and B indicate the boundaries for the metastable solution region with respect to crystallisation (a stable/unstable solution exists at lower/higher $S$) and the dashed curves in B and C highlight systems where $S$ is outside of this region. E shows the potential of mean force, $\phi$, as a function of the separation distance between Na$^+$ and Cl$^-$ in solution from Umbrella Sampling simulations which approach the limit of infinite dilution. The minima representing the contact ion pair (CIP) and solvent shared ion pair (SSIP) are highlighted. Error bars show the standard deviation in the data analysed from five, one nanosecond trajectories.}
    \label{fig:cluster}
\end{figure}

Before discussing NaCl crystallisation pathways from solution, we analyse the ionic species characterising a stable ($S \leq 1$) and metastable ($S>1$) solution at the steady state, \textit{i.e.} for timescales significantly shorter than the characteristic nucleation times. 
The solution achieves a quasi-equilibrium within such timescales; thus, ensemble averages pertaining to pre-critical species in solution can be directly computed via brute-force sampling.   
To this aim, a series of MD simulations were performed where NaCl molality ($b$) was 2--13.7~mol~kg$^{-1}$ and $S=0.5-3.7$.
We also performed simulations at and beyond the limit of solution stability ($S=4.1$\cite{jiang_nucleation_2019}) where $b=15$ and 16~mol~kg$^{-1}$ ($S=4.1$ and 4.3). 
Figure \ref{fig:Sscheme} in Supporting Information (SI) indicates how the superstauration levels targeted in this work compare to other computational studies and recent experiments from the literature.
Figure \ref{fig:cluster} shows how the non-ideality of the solution increases as the system is progressively brought further into the metastable region ($1<S<4.1$).
The average ion coordination number in Figure \ref{fig:cluster}~A increases from a value of around zero (\textit{i.e.,} indicating that ions are completely solvated by water molecules in the first coordination sphere) to 0.75 when $S=3.7$.
Across the same range of molality, the maximum coordination number increases from $1.1 \pm 0.2$ to $3.2 \pm 0.2$.
In the stable solution when $S=0.5$, the maximum coordination number is $0.7 \pm 0.2$, while at $S=4.3$, this is $3.6 \pm 0.2$ (albeit with a conservative definition of ion coordination; see Methods).
The increased ion-ion coordination, and partial dehydration across the metastable region, leads to clustering and the formation of extended ionic networks.
This observation is in very good agreement with results from recent experiments and simulations that showed the presence of extended Na--Cl ionic networks when $S$ is greater than 1.3. \cite{hwang_hydration_2021}

The extent to which clusters can grow is highlighted by the number-weighted cluster size probability distributions in Figure \ref{fig:cluster}~B, representing the probability of finding an ion within a cluster of size $N$. 
All distributions below the limit of solution stability show a rapid decay with respect to $N$; indeed, at the lowest concentration in the metastable region, an exponentially decaying function ($N \cdot p(N) \propto \exp (-1.8N)$) can be fitted to the data.
The complexity of the solution structure increases with concentration and large clusters emerge (the largest of which contain 8--18\% of the simulated ions at $S=3.7$).
We anticipate a system size dependence on the extent to which clusters can grow; both due to the number of ions available to form clusters in finite sized simulations\cite{wedekind2006finite,salvalaglio2016overcoming} and the size of the cluster domain with respect to the simulation cell geometry.
While Pulido Lamas \textit{et al.} \cite{pulido_lamas_homogeneous_2021} observed crystal nucleation in much larger brute force simulations when $S=3.8$ ($b=14$~mol~kg$^{-1}$ \textit{i.e.}, close to the limiting case of $S=3.7$ in the metastable region studied here), we did not observe spontaneous crystal nucleation (by monitoring $n(q6)$) in brute force simulations in the metastable region.
To verify this, we performed nine additional 300~ns MD simulations at $S=3.7$, where the initial solution configurations were generated independently, none of which crystallised.
Based on the rate for nucleation calculated in this work ($3.5 \times 10^{31}$~s$^{-1}$~m$^{-3}$, see below), the timescale for nucleation in this system is around 500~ns.


On crossing the spinodal, the cluster size distributions show markedly different behaviour (see the dashed curves in Figure \ref{fig:cluster}~B where $S=4.1$ and 4.3), with additional peaks in $N \cdot p(N)$ highlighting the spontaneous phase separation associated with large fluctuations of the local ion densities that occurs at these supersaturations.
The probability for the smallest clusters at the highest concentrations tends to decrease over time, concomitant with an increase in the size of the largest clusters. 
This indicates that, beyond the limit of solution stability, the timescale for phase separation is comparable with the length of our simulations and it is likely that the curves in Figure \ref{fig:cluster}~B for these cases would continue to evolve beyond the 50~ns used in our analyses.
At the end of these trajectories we also observed crystalline regions which develop spontaneously in the clusters, in agreement with simulations reported elsewhere. \cite{jiang_forward_2018,jiang_nucleation_2019,lanaro_birth_2016,pulido_lamas_homogeneous_2021,chakraborty_how_2013}

Further insight into the coordination environment in clusters can be gained by considering $p(n)$ which is correlated to the number of ions in the clusters with particularly high local ion density. 
The variable $n$, defined in the Methods section, represents the cubic root of the number of ions possessing a five-fold or higher ion coordination.
In the calculation of $n$, the function, $f$, describing coordination goes smoothly from one to zero over a distance 0.38--1~nm.
Figure \ref{fig:cluster}~C shows well defined monomodal distributions for $p(n)$ which shift to larger values of $n$ as the concentration of ions increases. 
Together with Figure \ref{fig:cluster}~A and B, this shows that in a relatively lean metastable solution the ionic species present before nucleation are predominantly dissociated ions.
As the system explores further into the metastable region, however, the large clusters which form contain regions of high ion density accounting for 0.1--0.3 of the total fraction of ions in the clusters.
Phase separation, which occurs spontaneously at the spinodal point, increases the likelihood of these dense regions, which is shown by a shifting to larger $n$ and the appearance of a shoulder in the peaks (see the dashed curves in Figure \ref{fig:cluster}~C) averaged over the entire MD simulation.

Finally, we investigated how ion clustering affects the dynamics of ions and water in solutions crossing the spinodal.
Figure \ref{fig:cluster}~D shows a slowing of the diffusion of all solution species as the concentration increases. However, the observed decay is monotonic, as also shown elsewhere, \cite{jiang_nucleation_2019} and the data does not provide clear evidence for a dynamic arresting of ions in the clusters.
Furthermore, as for the mean ionic coordination number, no discernible discontinuity in the diffusion of solution species is detected when crossing the spinodal, at least on the timescale of $\sim10^1$~ns.
The self-diffusion coefficients, $D$, for Na$^+$ and water can be accurately determined as $7 \times 10^{-6}S^{-0.66}$ and $1.5 \times 10^{-5}S^{-0.96}$~cm$^2$~ps$^{-1}$, respectively for the range of $S$ sampled here.
With only limited reduction to the dynamics of ions, it is not surprising that, in experiments, dehydration and reorganisation of ions in dense regions to the crystal coordination geometry occur over very short timescales; hence, special techniques such as rapid drying must be used to avoid NaCl crystallisation in the production of amorphous NaCl nanoparticles. \cite{amstad_production_2015}

The presence of extended ionic networks with relatively high and low regions of ion density raises questions about their potential role in nucleation. 
In the context of characterising nucleation pathways, it is also useful to consider whether ion association is thermodynamically favoured, as this could indicate nucleation via the prenucleation cluster (PNC) pathway. \cite{gebauer_stable_2008}
The mechanism for precipitation along the this pathway involves the association of PNCs during a microscopic liquid-liquid phase transition with a slowing of the solution dynamics in the proceeding liquid intermediate which can undergo further transformations to amorphous solids or crystals. \cite{gebauer_stable_2008,sebastiani_water_2017,avaro_stable_2020}
A significant debate has surrounded the role of PNCs in nucleation and even their existence. \cite{hu_thermodynamics_2012,smeets_classical_2017,henzler_supersaturated_2018,carino_thermodynamic-kinetic_2017,gebauer_classical_2018}
In the case of calcium carbonate, PNCs are described as dynamically ordered polymeric species. \cite{demichelis_stable_2011}
The thermodynamic stability of the PNCs, relative to free ions in solution, is ascribed to a strongly exergonic ion association reaction where $K_a^M$---the molar equilibrium ion association constant---is independent of cluster size \textit{i.e.,} for $(\mathrm{AX})_\mathrm{n} + (\mathrm{AX}) \rightleftharpoons (\mathrm{AX})_\mathrm{n+1}$, where A, X and AX represent anions, cations and ion pairs, respectively, the free energy change for the forward reaction is a constant value independent of n.
If $K_a^M>1$, then according to a multiple binding model, it was argued that a (meta)stable PNC population should exist in solution, regardless of the value of $S$. \cite{gebauer_classical_2018}

To calculate $K_a^M$, we performed Umbrella Sampling simulations to evaluate the potential of mean force for ion pairing, $\phi$, in a system that approached the limit of infinite dilution following the method described in detail in the work of Chialvo \textit{et al.} \cite{chialvo_na-cl-_1995}.
$\Delta \phi (r)$ is provided in Figure \ref{fig:cluster}~E and shows excellent agreement with studies elsewhere. \cite{zhang_dissociation_2020}
This indicates that the contact ion pair (CIP; $r_{\mathrm{Na-Cl}}=0.28$~nm) is less stable than the solvent shared ion pair (SSIP; $r_{\mathrm{Na-Cl}}=0.5$~nm).
Under these conditions, 
\begin{equation}
    K_a^M = \frac{4 \pi}{C_0} \int_{r_0}^{r_1} \exp \left({\frac{- \Delta \phi(r)}{k_{\mathrm{B}}T}} \right) r^2 \; dr
\end{equation}
where $r$ is the distance between ions, $k_{\mathrm{B}}$ is Boltzmann's constant and $T$ is temperature. 
$C_0$ is a constant that ensures mol$^{-1}$~dm$^{3}$ units for the equilibrium constant. 
The integration limit $r_0$ was defined as the minimum distance at which ions approach each other, while $r_1$ was 0.6~nm \textit{i.e.,} beyond the second maximum in $\Delta \phi (r)$.
We evaluated $\log_{10} (K_a^M)$ as $-0.41 \pm 0.04$ (mol$^{-1}$~dm$^{3}$).
This value is less negative than the prediction of $(K_a^M)$ by analytical models fitted to ion conductance measurements in solution; \cite{ho_electrical_1994} 
although, the measurements were collected at temperatures beyond 400~K, and thus a direct comparison here should be treated with caution.
A negative value of $\log_{10} (K_a^M)$ indicates thermodynamically favourable dissociation, even when all types of ion pair states described here are considered.

Perhaps a better indicator for ion association in solution is the equilibrium constant for the formation of a CIP from a SSIP, which can be calculated according to,
\begin{equation}
    K_\mathrm{CIP} = \frac{\int_{r_0}^{r_T} \exp \left({\frac{- \Delta \phi(r)}{k_{\mathrm{B}}T}} \right) r^2 \; dr}{\int_{r_T}^{r_1} \exp \left({\frac{- \Delta \phi(r)}{k_{\mathrm{B}}T}} \right) r^2 \; dr}
\end{equation}
here, $r_T$ is 0.35~nm \textit{i.e.,} the first maximum in $\Delta \phi (r)$ which represents the transition state between the two ion pair types.
$\log_{10} (K_\mathrm{CIP})$ was $-0.77 \pm 0.02$ in our calculations, indicating that, even in dilute solutions, a small fraction of ion pairs are contact ion pairs, as was already found using cluster population methods for determining the association constant. \cite{joung_molecular_2009}
Regardless of the choice of association constant, therefore, the equilibrium investigated will not lead to a thermodynamically stable population of clusters in the multiple binding model alluded to above. 
Furthermore, no stable cluster size population was detected throughout the metastable solution region, where the the probability distributions in Figure \ref{fig:cluster}~B indicate that monomers are the most stable species. 
In this regard, Ising lattice gas simulations showed that a broad distribution of cluster sizes in solution results from liquid-liquid demixing close to the critical point for the binodal curve for the liquid/liquid phase separation. \cite{wallace_microscopic_2013}
A dense liquid NaCl will likely be short lived; however, cluster populations from simulations could inform about the underlying physics which drives cluster formation, as was shown already in the case of \textit{e.g.,} disordered proteins. \cite{rana_phase_2021}

\subsection{Sampling Nucleation Events using Metadynamics}

\begin{figure}[H]
    \centering
    \includegraphics[width=1.0\linewidth]{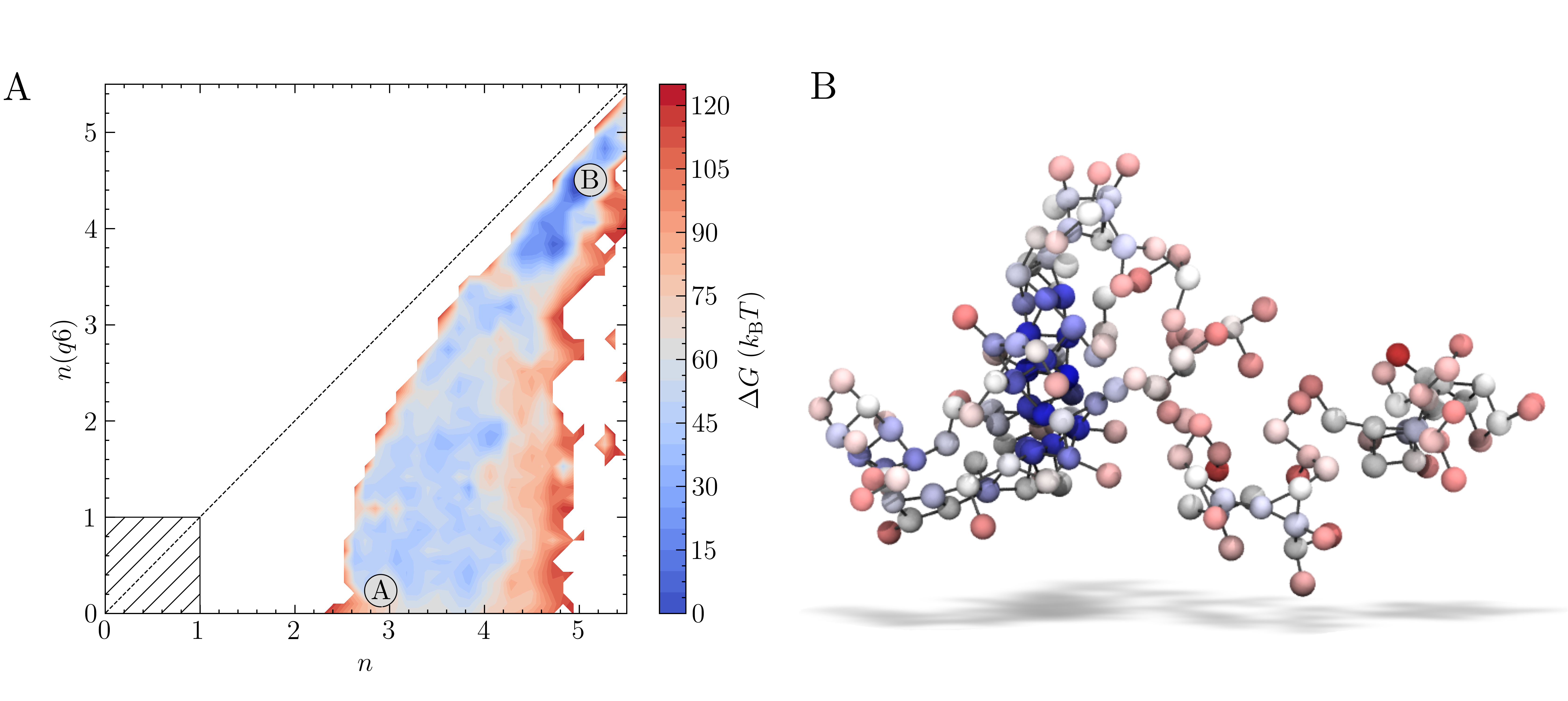}
    \caption{Pathways from solutions to crystals from biased sampling calculations when $S=3.7$. A provides relative Gibbs free energies, $\Delta G$, as a function of the reaction space variables $n$ and $n(q6)$ characterising the size of dense and crystalline NaCl regions, respectively. A and B marked on the plot indicate the solution and crystal states.
    B shows a liquid-like NaCl cluster observed in MD simulations where $S=3.7$. The hatched region indicates the range of $n$ and $n(q6)$ where less than one ion displays five-fold ion coordination or a rock salt crystal geometry, respectively. The grey lines emphasize first-sphere connections between the ion spheres which are coloured red $(c=0.6) \rightarrow{\mathrm{blue}\; (c=6)}$ according to their value of $c \;$: the coordination environment which informs about the relative ionic density in the cluster and is used to calculate $n$ (see Methods).}
    \label{fig:fes}
\end{figure}

Well-tempered metadynamics (WTmetaD) simulations were initiated both from homogeneous supersaturated solutions and from crystals of various sizes immersed in solution (see Methods) to analyse the pathways for crystallisation when $S=3.7$ \textit{i.e.,} a supersaturation far into the metastable region. 
10~\textmu{s} total simulation time was generated from 20 individual simulations. 
The added bias potential ($V(n^{sph})$) in WTmetaD enhances the sampling of the high ion density regions in the clusters within a spherical volume in the simulation cell (see Methods).
Localising the enhancement of ion density fluctuations triggered by the bias is necessary at high $S$ to limit the number of crystal nucleation events leading to the formation of structures stabilised by periodic boundary conditions. 
No constraint was applied to the position of the nuclei emerging from solution, however, and these can translate outside of the spherical volume.
We therefore post-process the trajectories (see Methods) to calculate the probability density of states in the space $n,n(q6)$.
Calculating the forces on atoms from the derivatives of the bias in $n(q6)$ is computationally very expensive, which makes biasing this CV in simulations intractable for the adopted system size and simulation times. 
This offers an advantage in that by introducing bias only to increase the fluctuations in local density, we avoid forcing the system to adopt any particular crystalline structure.
However, since fluctuations in the CVs defining the energy landscape were not directly enhanced during the sampling, some regions of the reaction coordinate are less well converged than others; indeed, out of the 20 WTmetaD simulations, four failed to visit both the solution and crystalline states (and transition states between them) and were therefore discarded in subsequent analyses.

The free energy landscape in Figure \ref{fig:fes}~A indicates that a wide range of states with minimal crystalline order corresponding to small values of $n(q6)$ are accessible.
The minimum accessible $n$ is around 2.5 when $n(q6) = 0$, in agreement with our unbiased simulations. 
A value of $n=2.5$ indicates that around 15 ions in solution have high ion coordination and form partially dehydrated ion clusters.
The configuration in Figure \ref{fig:fes}~B provides an example of an ion cluster which emerged in MD simulations of NaCl(aq) solutions when $S=3.7$; the colours for the ions indicate their relative coordination number for the case where $n=2.8, \, n(q6)=0$. Our MD simulations indicated that the average number of ions in the largest cluster at $S=3.7$ was $100 \pm 40$.
It is clear that non-uniform chemical ordering occurs in the ion clusters: high density regions, shown by the blue spheres in Figure \ref{fig:fes}~B, are surrounded by extended ionic networks where ion coordination is much lower, and where the geometry and connectivity of the cluster evolves rapidly during the simulations.
The energy landscape generated using WTmetaD indicates that $n$ states beyond the maximum found in MD simulations, up to $n \approx 4$, are accessible in the long time limit.
The molality of ions in the high density regions when $S=3.7$, therefore, is around 0.6~mol~kg$^{-1}$, but this can increase by several mol~kg$^{-1}$ as the system explores $n$.

As the degree of crystalline order increases, the accessible region in $n$ narrows significantly.
Up to $n(q6) \approx 2$, the range of accessible $n$ states is largely independent of $n(q6)$, but it is around this point that the accessible states converge towards a limiting case for crystallisation on the diagonal, where the size of the dense and crystalline regions increase simultaneously.
The crystal basin is observed beyond $n=5$ and $n(q6)=4.5$.
Although the definition of $n(q6)$ was able to correctly identify crystal ions at rock salt planar surfaces, the rough crystal nanoparticle surfaces found in solution limits the accessibility of states on the diagonal to systems which contain particularly large crystals.
From the minimum corresponding to the metastable solution and labelled `A', lowest energy pathways to the crystal, `B', were calculated using nudged elastic band (NEB) calculations. \cite{henkelman_improved_2000}
While these correctly identified the crystal as the most stable state, the profiles were noisy and no clear transition state for crystallisation was identifiable with confidence. 
Nevertheless, WTmetaD was instrumental to obtain an extensive sampling of configurations spanning the entire reaction coordinate space which are inaccessible to sub-\textmu{s} seeded MD or solution simulations.


\subsection{Nucleation Pathways in Unbiased Simulations}

\begin{figure*}[th]
    \centering
    \includegraphics[width=0.8\linewidth]{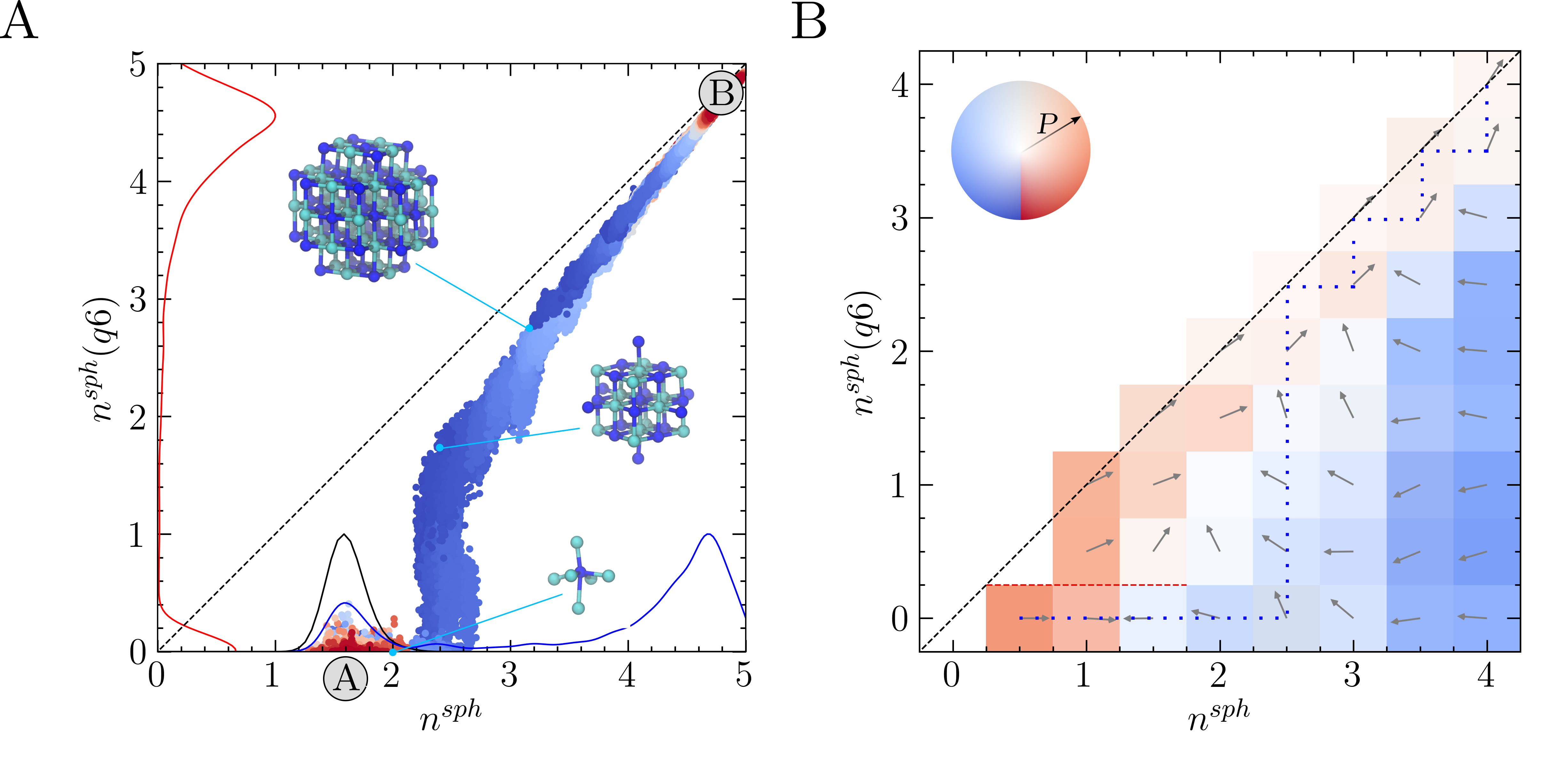}
    \caption{A. Seeded MD simulations initiated from crystal seeds with varying sizes; the colour scale here indicates the time in the simulation as blue $(t=0)$ $ \rightarrow{}$red ($t=300$~ns). Also provided by the solid lines are the probability densities in the reaction space variables $n^{sph}$ (blue) and $n^{sph}(q6)$ (red) calculated using all of the data from seeded MD and the probability density from MD simulations of the solution phase (black). Inset are configurations for three crystal seeds around $t=0$, with blue and cyan spheres indicating Na$^+$ and Cl$^-$ ions, respectively. The initial values of $n^{sph},n^{sph}(q6)$ for these seeds are highlighted by the light blue points. B The mean flux of trajectories on a discretised $n^{sph},n^{sph}(q6)$ reaction coordinate space sampled using swarms of MD simulations. The mean path for trajectories from the each bin are shown by the arrows and colours, with the opacity of the colour indicating the probability of transition to adjacent bins occurring on the timescale of the simulations. The red dashed line highlights the case where zero probability of transition to adjacent bins occurs.}
    \label{fig:discrete}
\end{figure*}

\paragraph{MD Simulations}
To obtain a quantitative estimation of transition paths, we took advantage of the extensive sampling of configurations obtained from WTmetaD to initialise a comprehensive set of unbiased MD trajectories covering the entire reaction coordinate space. 
For the analysis of the unbiased trajectories, we focus our discussion on the $n^{sph},n^{sph}(q6)$ reaction coordinate space: a localisation of the $n,n(q6)$ CVs in the centre of the simulation cell.
Considering a subspace of the simulation cell reduces the noise associated with diffusion of the system in the reaction coordinate and, because the crystal seeds discussed below were anchored to the centre of this subspace, no translation of the nanoparticles occurs during the simulations, which would otherwise lead to inaccuracies in the relative probability density of states. 
Nonetheless, the general features we describe below were apparent regardless of this choice in the reaction coordinate.
Our MD simulations indicated that the mean $n^{sph}$ from extended simulations of the homogeneous solution phase is $1.61 \pm 0.13$ and the normalised probability density for this is shown in Figure \ref{fig:discrete}~B. 

Two approaches were taken to investigate diffusive fluxes in the reaction coordinate space.
In the first of these, 10 crystalline seeds of various sizes were simulated in solution for 300~ns and their growth or dissolution was monitored by tracing trajectories in $n^{sph},n^{sph}(q6)$.
Figure \ref{fig:fes}~A shows that, depending on the size of the initial crystal seed, systems consistently relax to either the solution (A) or crystal states (B). Overlapping multiple seeded trajectories provides a representation of the path between these states in the two-dimensional reaction coordinate space $n^{sph},n^{sph}(q6)$.
While the smallest and largest crystal seeds dissolved and grew, respectively, seeds with diameters in the range 0.8--1~nm relaxed to states which were opposite to this general trend, as shown in Supporting Information (SI) Figure~\ref{fig:mdpaths}.
A seed containing 19 ions where $n^{sph}(q6)=1.3$, and which was highly faceted, grew in solution; whereas, a seed containing 27 ions with perfectly planar surfaces, and where $n^{sph}(q6)=1.5$ (since in both seeds only one ion has complete rock salt coordination geometry), dissolved.
Ion coordination between seeds and other liquid-like clusters occurred readily at the beginning of the simulations and ion exchange with the surrounding ionic networks ultimately consumed these less dense regions of the clusters when the seeds were post-critical. 
The probability densities shown in Figure \ref{fig:fes}~A indicate that the pathway to bulk crystals involves increasing the density in ion clusters in a first step to crystal nucleation: a departure from the mean $n^{sph}$ in solutions at the steady state (see the black curve) occurs before crystalline order emerges, when $n^{sph}(q6)>0$, in a second step.
This is particularly significant as it indicates that crystallisation occurs in regions where relatively high levels of ion-ion coordination, and therefore low levels ion solvation, are observed compared to the homogeneous solution state.
The minimum in the probability density in $n^{sph}(q6)$ is 0.5--0.6; here, $n=2$ indicating at least eight ions---on average---in the subspace with an ion rich first coordination sphere. 
The rate limiting step to crystallisation, therefore, is the reordering of ions in the dense liquid domains to a geometry which begins to resemble that of the crystalline phase.

Further support for the two-step pathway to phase separation is gained from our second approach to investigate diffusive flux in the reaction coordinate space.
Here, the $n^{sph},n^{sph}(q6)$ space was discretised into $0.5 \times 0.5$ bins.
Swarms of MD simulations were initiated in each bin where the initial configurations for the simulations were extracted from MD simulations of solutions and from the WTmetaD simulations already described.
Initial configurations were selected randomly, ensuring that no two initial starting points were correlated.
At least 40 independent 2.5~ns simulations were initiated in each bin which was increased to $\sim 100$ simulations in regions of the configuration space where diffusion is particularly slow.
Figure \ref{fig:discrete}~B shows the direction of flux for the average trajectory initialised in each bin.
Using this coarse-grained representation of the flux of configurations in reaction coordinate space, the two-step nucleation pathway spontaneously emerges, and is clearly apparent. 
This is highlighted by the most probable pathway which propagates the system beyond previously visited states in the solution phase to the crystal based on the transitions between adjacent bins, as shown by the blue dotted line in Figure \ref{fig:discrete}~B. 
Here, $n^{sph}(q6)$ initially increases when $n^{sph}=2.5$ until the system approaches and follows the diagonal beyond $n^{sph}=2.5,n^{sph}(q6)=2.5$, in close agreement with the pathway identified in Figure \ref{fig:discrete}~A.
Furthermore, the qualitative features here are in good agreement with the free energy landscape provided in Figure \ref{fig:fes}~A.

Systems where $n^{sph}(q6)=0$ and $n^{sph}$ is particularly small diffuse rapidly to states where the ion density in clusters increases and there was zero probability of transition from bins where $n^{sph}<2$ and $n^{sph}(q6)=0$ to those where $n^{sph}(q6)>0$ during the 2.5~ns timescale.
When $n^{sph}>2.5$ and $n^{sph}(q6)=0$, the system tends to diffuse towards $n^{sph}=1.5-2.5$, with the probability of transitions (highlighted by the opacity of the colours in the figure) increasing with $n^{sph}$.
This provides confidence to the qualitative assessment regarding the accessibility of states with increased ion coordination relative to the most probable state for a homogeneous solution.
It is notable too that the direction of the mean path when $n^{sph}=1,n^{sph}(q6)=1$ indicates concerted growth in the reaction space variables.
A pathway along the diagonal from $n^{sph}=1,n^{sph}(q6)=1$ could be consistent with single-step nucleation occurring via the addition of single monomers to growing crystal nuclei; although technically, these nuclei could still reside within a liquid-like cluster.
Therefore, while the most likely pathway under these conditions appears consistent with two-step nucleation, multiple, coexisting pathways may contribute to phase separation.

\paragraph{Markov State Model}
To gain quantitative information about the nucleation pathways spontaneously emerging from unbiased MD, we used the data gathered  from all unbiased simulations, amounting to a total simulation time of 20~\textmu{s}, to construct a Markov State Model (MSM) for the nucleation process. 
Here, the reaction coordinate was partitioned into approximately 100 discrete states which are shown in Figure \ref{fig:msm-sph-si}~A and \ref{fig:msm-box}~A in SI. The MSM provides independent insight into the free energies associated with the nucleation process, the implied timescales associated with the relaxation of the system, as well as the information necessary to characterise the nucleation process using tools from discrete transition path theory.  

\begin{figure}[H]
    \centering
    \includegraphics[width=1\linewidth]{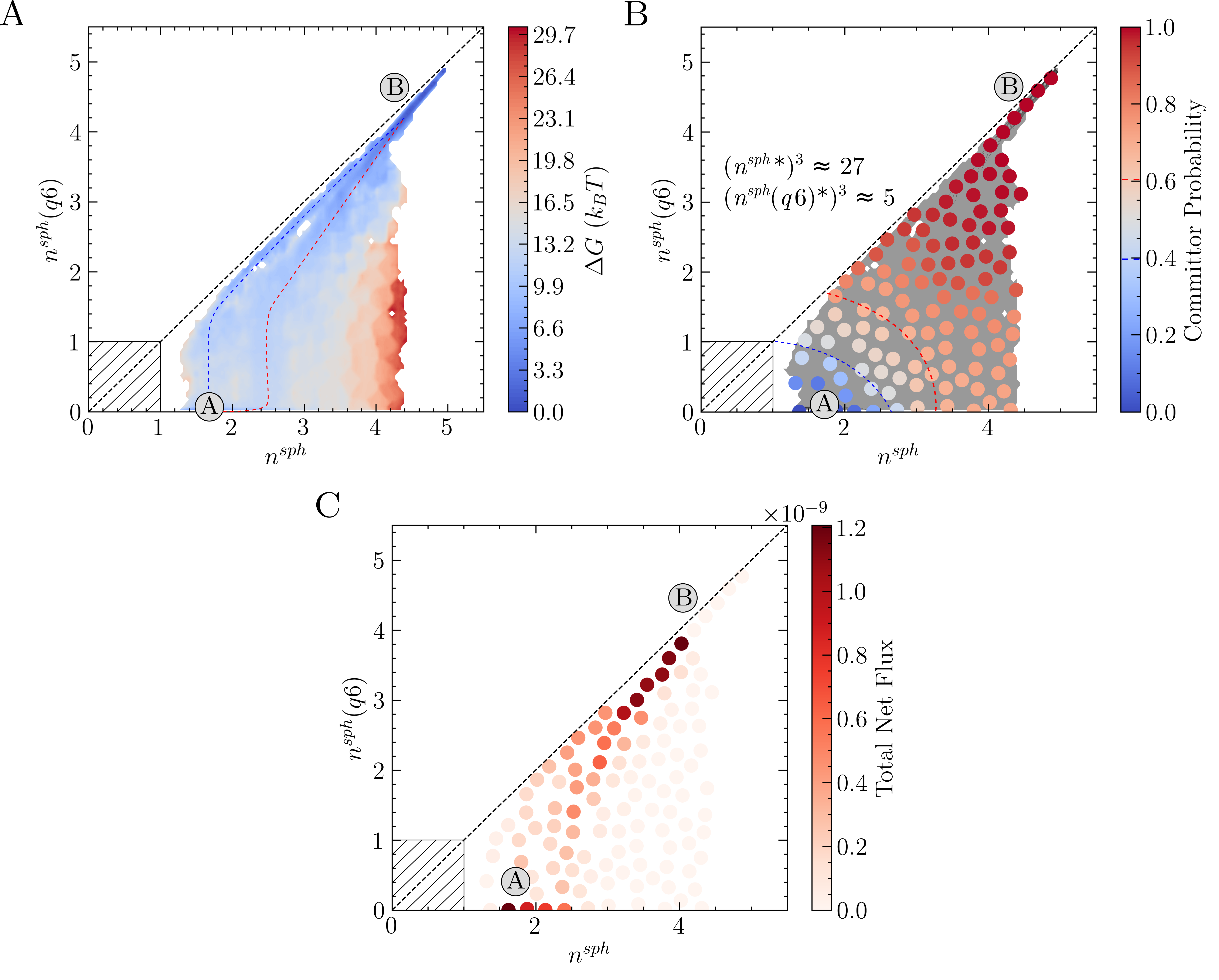}
    \caption{ 
    A, B and C provide the free energy landscape, committor probabilities and total net flux between partitioned states, respectively, in the $n^{sph},n^{sph}(q6)$ reaction coordinate space calculated from the MSM described in the text. The hatched regions indicate the range of $n^{sph}$ and $n^{sph}(q6)$ where less than one ion displays five-fold ion coordination or a rock salt crystal geometry, respectively. The blue and red dashed lines in A provide a guide to the eye for the lowest energy one- and two-step nucleation pathways, respectively. The blue and red dashed lines in B indicate the approximate range of the transition state ensemble of states where the committor probability is $0.4-0.6$ (see key). 
    }
    \label{fig:msm}
\end{figure}

The free energy landscape in Figure \ref{fig:msm}~A shows two basins for the solution and crystal states (labelled A and B, respectively) and again confirms the wide distribution of accessible $n^{sph}$ states corresponding to clusters with negligible crystalline order.
Following the two-step pathway already identified in Figure \ref{fig:discrete}, the barrier to crystallisation is approximately 6~$k_\mathrm{B}T$, in reasonable agreement with the barrier determined elsewhere by interpolating the data from umbrella sampling simulations and mean fist passage time in forward flux sampling. \cite{jiang_nucleation_2019}
An alternative pathway exists from the solution basin to $n^{sph}=n^{sph}(q6)=1$, and which subsequently follows closely to the diagonal towards the crystal basin, is also evident in the energy landscape; this is captured due to the finer partitioning of states here compared to the coarse approach taken in Figure~\ref{fig:discrete}~A. 
The barrier for crystal nucleation along this pathway is around 10~$k_\mathrm{B}T$, which further supports the notion that multiple pathways for nucleation are available to the system far into the metastable region.
The energy landscape in Figure \ref{fig:msm}~A is compatible with those theorised using a general framework to describe two-step nucleation pathways (using CNT concepts) recently proposed by Kashchiev \cite{kashchiev_classical_2020}.
Both one- and two-step pathways to crystal nucleation are predicted by this framework when the maximum in the free energy landscape---characterised by cluster size and crystalline order---is associated with a large amorphous cluster (see Ref. \citenum{kashchiev_classical_2020}, Figure~3a).
The distinction between the pathways identified in Figure \ref{fig:msm}~A tends to fade when the free energy in the space $n,n(q6)$ is analysed (see Figure \ref{fig:msm-box}~C in SI), due to the noise associated with $n$ fluctuations in the entire simulation cell.

A committor analysis was performed to calculate the probability for partitioned states (associated with the MSM) in the reaction coordinate space to relax to either the solution or crystal basins. 
This analysis (see Figures \ref{fig:msm}~B and \ref{fig:msm-box}~D) revealed that the transition state ensemble, identified here by states with committor probability values of 0.4--0.6, spans a wide range of cluster configurations including both fully ordered and completely disordered ones. 
Indeed, even for the largest clusters with no apparent crystalline order (\textit{i.e.,} when $n^{sph}$ is beyond 2 and $n^{sph}(q6)$ is zero), the committor is larger than 0.5, indicating their post-critical nature \textit{i.e.,} they display a greater likelihood to crystallise than to dissolve. 
This observation is evidence for the coexistence of multiple pathways from the solution to crystal states and introduces the necessity of generalising the concept of a critical nucleus, to that of a \textit{critical ensemble of nuclei}. 
Our simulations indicate that, within the critical ensemble, the average cluster contains around 27 ions with five-fold coordination or more, 5 of which have rock salt crystal coordination geometries.

The MSM provides quantitative estimates for the kinetics associated with state-to-state dynamics. By interpreting the slowest implied timescale as the time for relaxation of the system between states A and B (Figures \ref{fig:msm}~A-C), we obtain a nucleation rate of $3.5 \times 10^{31}~\mathrm{m}^{-3}~\mathrm{s}^{-1}$ (95\% confidence interval: $2.7 \times 10^{31}$--$5.1 \times 10^{31}~\mathrm{m}^{-3}~\mathrm{s}^{-1}$).
Alternatively, the nucleation rate has been computed from the mean first passage time between two metastable states identified using the robust Perron cluster analysis\cite{roblitz2013fuzzy} as $6.0\times 10^{31}~\mathrm{m}^{-3}~\mathrm{s}^{-1}$. Both estimates are in excellent agreement with the rates determined elsewhere using alternative methods. \cite{jiang_nucleation_2019,pulido_lamas_homogeneous_2021}. 

Finally, we project the total current of reactive trajectories in state space, defined for state $i$ as $\frac{1}{2}\sum_j\lvert{F_{ij}}\rvert$, where $F_{ij}$ is the current of reactive trajectories between states $i$ and $j$.
This provides a quantitative analysis of the transition mechanism between states A and B, which further supports the emergent picture for phase separation described so far. 
In particular, analysis of the high current trace between states, shown in Figure \ref{fig:msm}~C, indicates the presence of a main reaction channel corresponding to the most probable pathway identified by seeded simulations (Figure \ref{fig:discrete}A), and by analysing the flux of configurations directly obtained from swarms of short MD simulations (Figure \ref{fig:discrete}B). Interestingly, a secondary channel, closer to the definition of a one-step mechanism and associated to lower values of the net flux can be identified, quantitatively supporting the co-existence of multiple pathways from solution to crystals in the metastable region. 

Jiang \textit{et al.} \cite{jiang_nucleation_2019} sampled similar reaction coordinate spaces using forward flux sampling simulations and brute force simulations. They find that the nucleation pathway below the spinodal (when $S=2.7$ and 3.2) occurs in a single-step through correlated fluctuations in the density and crystallinity of emerging clusters.
Beyond the spinodal (when $S=4.3$), the phase separation mechanism was consistent with two-step nucleation, where a wide distribution in the size of amorphous clusters was observed before the formation of crystalline regions (as we also observed in MD simulations at this molality). 
It is possible then that, on moving the system far into the metastable region, there is a transition in the most probable pathway from one- to two-step nucleation and this may occur in the window $S=3.2-3.7$. Multiple accessible pathways to crystals allows for greater control of the nucleation process, as was shown mechanistically for the case of urea. \cite{salvalaglio_urea_2015,salvalaglio2015molecular}

\section{Conclusions}

\begin{figure*}[th]
    \centering
    \includegraphics[width=0.75\linewidth]{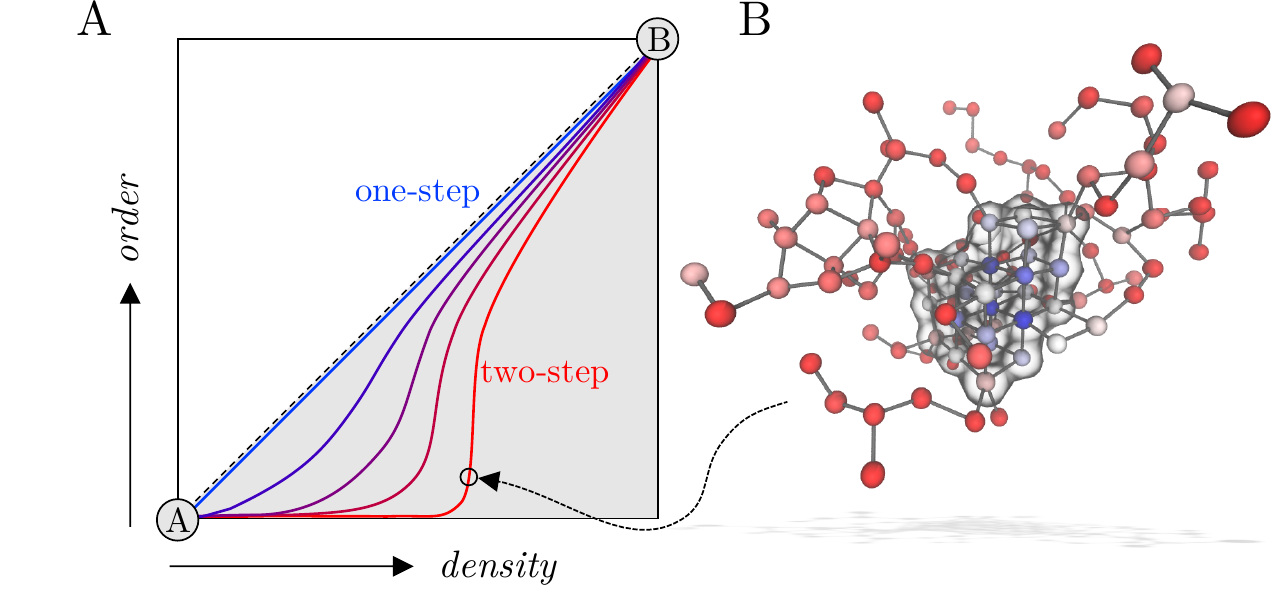}
    \caption{A shows a schematic of the two-dimensional reaction coordinate sampled in this work. Crystallisation pathways from a metastable solution phase (A) to states where a NaCl crystal is in equilibrium with a saturated solution (B) are highlighted. A pathway consistent with one-step nucleation is provided by the blue path on the diagonal. Paths which deviate from the diagonal indicate inaccuracies associated with the capillary approximation of CNT, while paths which go via extended, high density amorphous clusters (in which crystallisation subsequently occurs) suggest a two-step nucleation mechanism. B provides a snapshot of a pre-critical cluster along the two-step pathway containing around 130 ions and where $n^{sph}=2.4$, $n^{sph}(q6)=0.8$. Ions are coloured according to their local value of $q6$ in the range 0 (red) to 0.6 (blue), where a value of 1 would indicate a perfect crystal coordination geometry for ions. The transparent surface highlights those ions which begin to adopt a pseudo-crystal rock salt lattice structure with (111) crystallographic orientation in the projection shown.}
    \label{fig:nuc-scheme}
\end{figure*}

In this work, homogeneous NaCl nucleation mechanisms from solutions far into the metastable region were analysed in detail. 
To this aim, an extensive set of molecular trajectories obtained using different computational methods, including biased and unbiased dynamics were generated. 
In our analyses we explicitly considered nucleation as a process evolving along a multidimensional reaction coordinate space where both the size for ion dense regions and crystalline order in emerging nuclei evolve along pathways connecting metastable solutions and crystalline phases.
An analysis of the cluster size populations in this work shows that ion clusters are apparent in solutions at all concentrations in the metastable region at the steady state (before nucleation), in line with experiments;\cite{hwang_hydration_2021} although no special thermodynamic status can be attributed to these species, as was suggested in other mineralising systems. \cite{gebauer_classical_2018}
In metastable solutions, the clusters emerge through density fluctuations and are unstable with respect to dissociated ions; however, beyond the spinodal, aggregation leads to extended amorphous clusters during a spontaneous phase separation.
Both the average size for the largest clusters and the size of high ion density regions in the clusters (where there is a significant level of dehydration) increase as the system explores further into the metastable solution region.

Inspired by the recent approach of Kashchiev\cite{kashchiev_classical_2020} to provide a thermodynamic and kinetic framework for two-step nucleation, we analysed free energy landscapes characterising the size of the dense cluster regions and the degree of crystalline order in clusters (see the schematic in Figure~\ref{fig:nuc-scheme}~A). These show that multiple nucleation pathways are available to the system when \textit{b}(NaCl(aq)) = 13.7~mol~kg$^{-1}$ and $S=3.7$, and we identified the lowest energy one- and two-step routes to crystals from the range of possible pathways, beginning from the solution phase (see Figure \ref{fig:nuc-scheme}~A).
Two-step nucleation is the most probable of these pathways to phase separation under the conditions studied, and a wide range of states are accessible where clusters which contain regions of high ion density (compared to lower ion density clusters on average in solution) are completely amorphous; indeed, the committor computed via the construction of a MSM showed that a critical ensemble of nuclei exist, and that even large amorphous clusters tend to relax to a crystal rather than dissolving.
The activation energy for crystallisation along the one-step pathway we identify is greater than in the case of two-step nucleation.
It is important to note that nucleation in a single step, defined in terms of the reaction space variables $n$ and $n(q6)$, could occur by direct association of dissociated ions into a crystal nucleus or by concerted increases to the density and crystalline order of ions already present in liquid-like clusters in solution.
From an experimental perspective, distinguishing these types of one-step nucleation mechanisms would be challenging, particularly at large values of $S$.
Figure~\ref{fig:nuc-scheme}~B shows a pre-critical cluster along the two step pathway which contained around 100 ions. 
Crystal nucleation occurs in the high density region of the cluster shown by the blue spheres and highlighted by the transparent surface in the image. The kinetics of nucleation computed here, explicitly acknowledging the complexity of the configuration space, showed excellent agreement with those calculated elsewhere. \cite{jiang_nucleation_2019,pulido_lamas_homogeneous_2021}

Given the finding that crystal nucleation in NaCl(aq) solutions occurs via a single-step pathway when \textit{b}(NaCl(aq)) = 12~mol~kg$^{-1}$, \cite{jiang_nucleation_2019} we hypothesise that the transition from predominantly one- to two-step nucleation occurs in the metastable region at $S=3.2-3.7$; however, we predict that large amorphous clusters with high density ion regions could still provide pathways to crystals within this range of supersaturation.
In other mineralising systems, amorphous mineral phases are observed during two-step nucleation from solutions. \cite{faatz_amorphous_2004,wolf_early_2008,avaro_stable_2020,smeets_classical_2017}
On the other hand, amorphous NaCl phases are only stabilised using special preparation techniques. \cite{amstad_production_2015}
The difference here may simply be due to undersaturation of the solutions with respect to a dense liquid (if one exists) or amorphous solid NaCl phase; nonetheless, this scenario could still lead to a two-step nucleation pathway, \cite{kashchiev_classical_2020} where nascent amorphous clusters rapidly transform to crystals, if the energy barrier to forming the amorphous state in the first step is lower than the barrier to crystal formation in one-step. It is possible that kinetic stabilisation of the amorphous phase occurs in other systems, where the energetic barriers associated with translating bulkier anions to their crystalline geometries results in longer-lived amorphous precipitates.

Recently, we showed that surfaces catalyse the formation of large liquid-like NaCl clusters in solutions where the chemical potential is held constant in C$\mu$MD simulations\cite{finney_electrochemistry_2021,finney2021bridging}.
Both the size and density of ions in the clusters were increased in the presence of graphite surfaces. 
Within the context of CNT, interfaces reduce the surface energies of emerging nuclei and therefore the thermodynamic barriers to nucleation. Given the pivotal role of dense clusters in the two-step nucleation pathway, it may also be the case that surfaces catalyse a change in the nucleation mechanism from predominantly one-step to two-step in the metastable region, at lower supersaturations than those analysed here. 
This observation opens up new avenues to experimentally validate the mechanistic picture emerging from simulations, and to exploit surfaces to control nucleation.


\begin{acknowledgement}

The authors acknowledge funding from an EPSRC Programme Grant (Grant EP/R018820/1) which funds the Crystallization in the Real World consortium. The authors acknowledge the use of the UCL Myriad High Throughput Computing Facility (Myriad@UCL), and associated support services, in the completion of this work.

\end{acknowledgement}

\begin{suppinfo}
Additional figures are included in the associated supporting information.
PLUMED input files used in this work are available via PLUMED-NEST (https://www.plumed-nest.org \cite{the_plumed_consortium_promoting_2019}), the public repository for the PLUMED consortium, using the project ID: plumID:21.044.

\end{suppinfo}

\bibliography{NaCl_nucleation}

\clearpage
\pagenumbering{arabic}
\appendix
\setcounter{figure}{0} 
\setcounter{table}{0} 
\setcounter{equation}{0} 
\renewcommand\thefigure{S\arabic{figure}}
\renewcommand\thetable{S\arabic{table}}

\newpage
\onecolumn
{
\centering
~~
\section{Multiple Pathways in NaCl Homogeneous Crystal Nucleation}

\vspace{0.5 cm}
{\Large \underline{\underline{Supporting Information}}} 
\vspace{1 cm}

{\large Aaron R. Finney and Matteo Salvalaglio}
\vspace{0.5 cm}

\textit{Thomas Young Centre and Department of Chemical Engineering, University College London, London WC1E~7JE, United Kingdom}
\vspace{0.5 cm}

E-mail: a.finney@ucl.ac.uk; m.salvalaglio@ucl.ac.uk

}
\vspace{1 cm}
\etocdepthtag.toc{mtappendix}
\etocsettagdepth{mtchapter}{none}
\etocsettagdepth{mtappendix}{paragraph}
\etocsetnexttocdepth {paragraph}

~~

\section{Additional Figures}
\addcontentsline{toc}{section}{Additional Figures}
\label{sec:addfigures}

\begin{figure*}[th]
    \centering
    \includegraphics[width=0.65\linewidth]{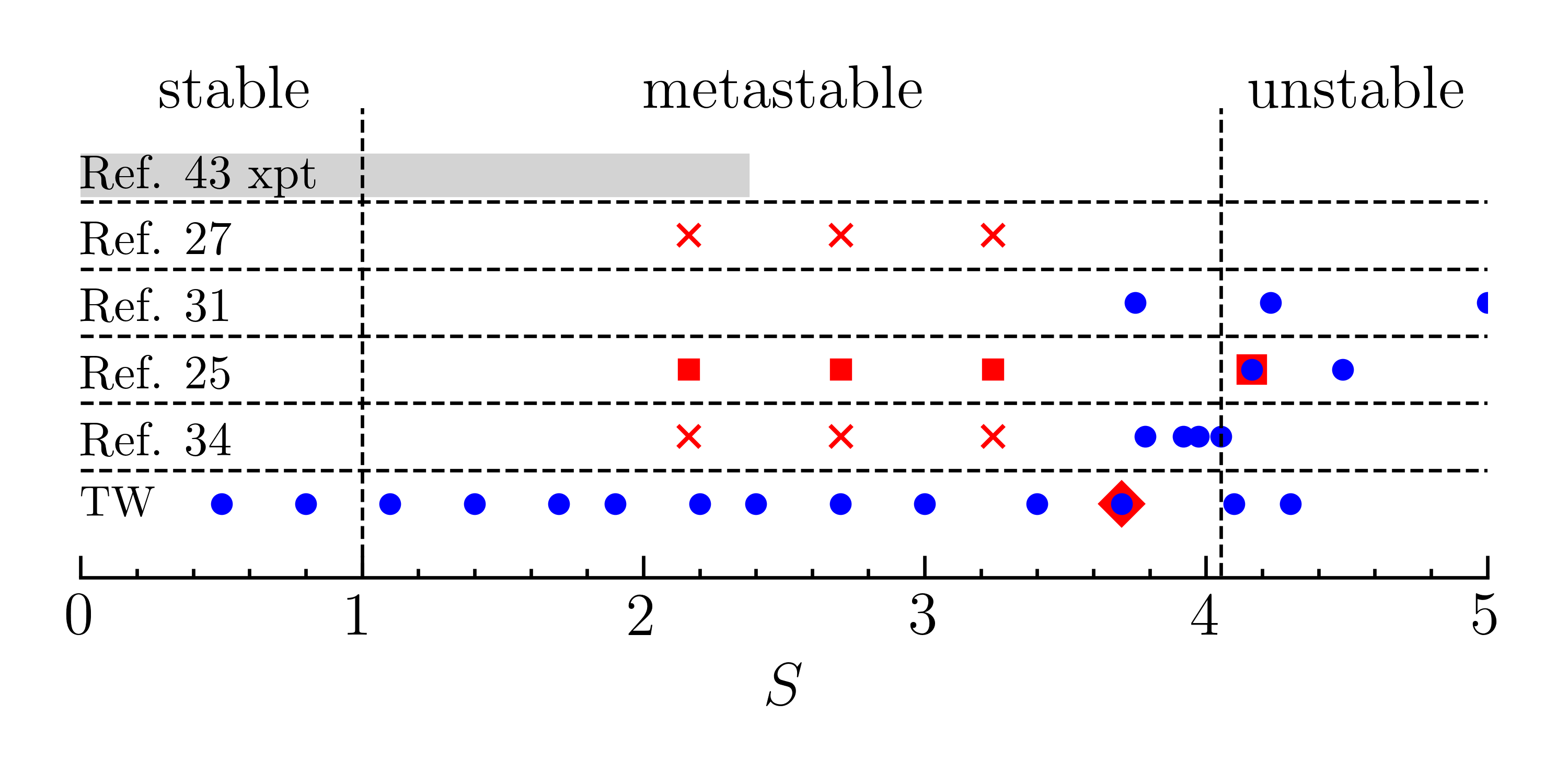}
    \caption{Supersaturation ($S$) levels adopted in recent simulations (using the force field from Ref. \citenum{joung_determination_2008} at room temperature) and experiments, as indicated by the references to works in the main paper. ‘TW’ refers to the supersaturations simulated in this work. Circles, crosses and squares indicate brute force MD, seeded MD and forward flux sampling simulations, respectively. The diamond indicates the combination of seeded MD and metadynamics adopted in this work to study nucleation at $S=3.7$. The grey bar indicates the range of $S$ measured in experiments of levitated droplets of NaCl(aq) solutions where liquid-like clusters were detected at the high end of $S$. }
    \label{fig:Sscheme}
\end{figure*}

\begin{figure*}[th]
    \centering
    \includegraphics[width=0.8\linewidth]{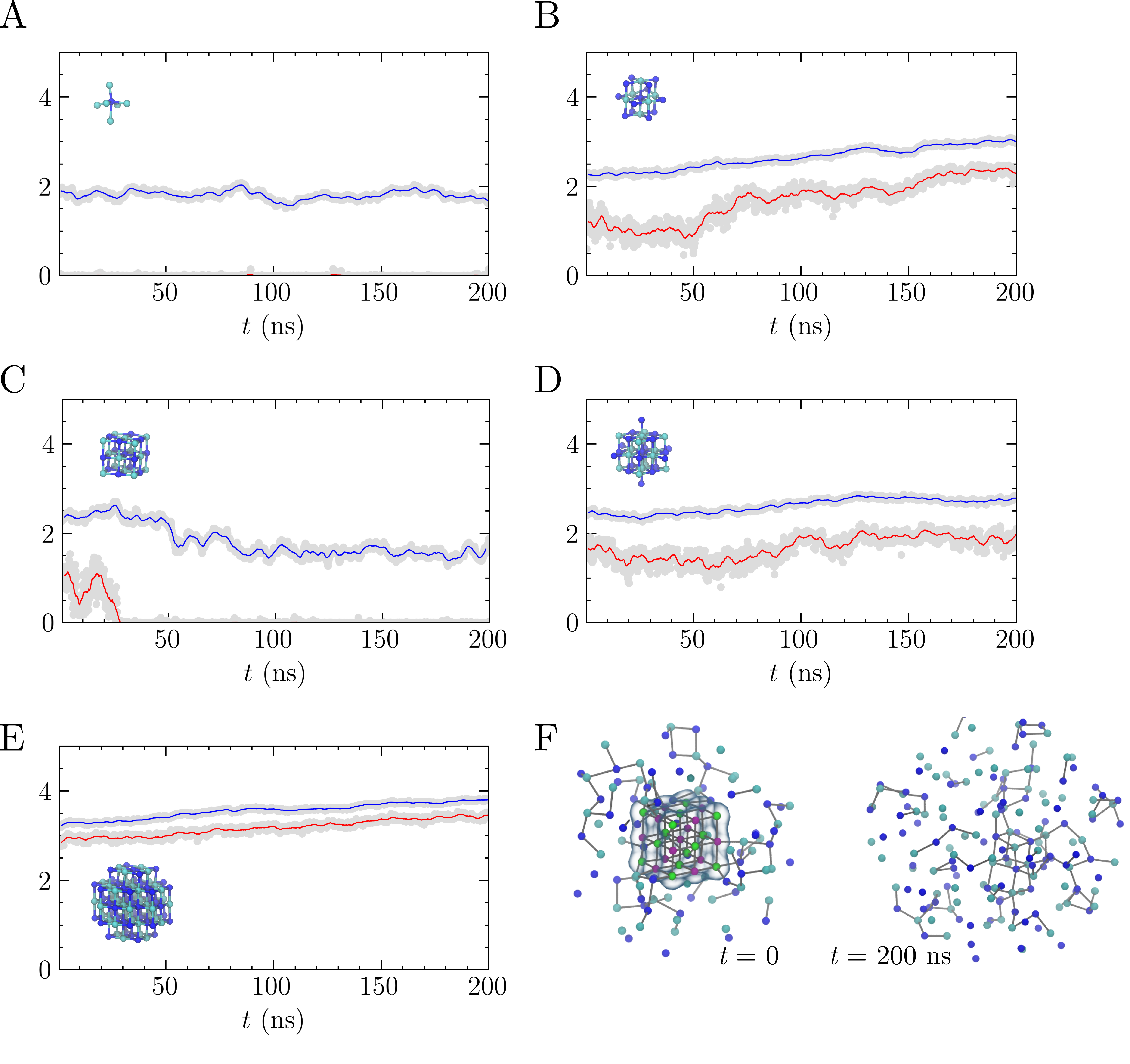}
    \caption{Simulations of crystalline seeds in solution where the diameter of the seeds was $0.6-1.4$~nm in 0.2~nm increments for A--E. The time series for $n^{sph}$ (blue) and $n^{sph}(q6)$ are provided. Inset are the initial configurations for the seeds which were immersed into solution. F shows example configurations for the species in solution from the simulation in C at $t=0$ and 200~ns. Blue and cyan spheres represent Na$^+$ and Cl$^-$, respectively, and the first sphere coordination between ions is highlighted by the grey lines. The ions in the initial crystalline seed in F are highlighted by the purple (Na$^+$) and green (Cl$^-$) spheres.}
    \label{fig:mdpaths}
\end{figure*}

\begin{figure*}[th]
    \centering
    \includegraphics[width=1\linewidth]{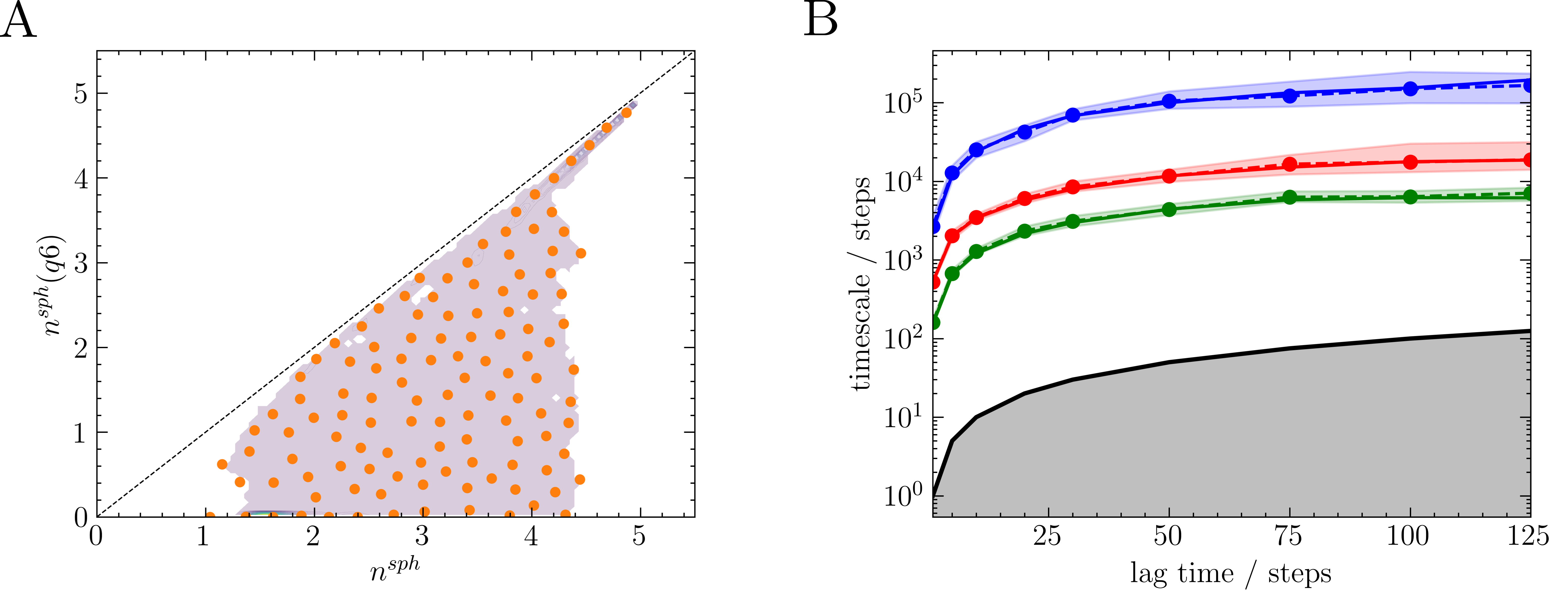}
    \caption{A: Points show the partitioning of the reaction coordinate into $\sim 100$ states used in the Markov State Model described in the main text. B provides the timescale for different lag times between the states.}
    \label{fig:msm-sph-si}
\end{figure*}

\begin{figure*}[th]
    \centering
    \includegraphics[width=1\linewidth]{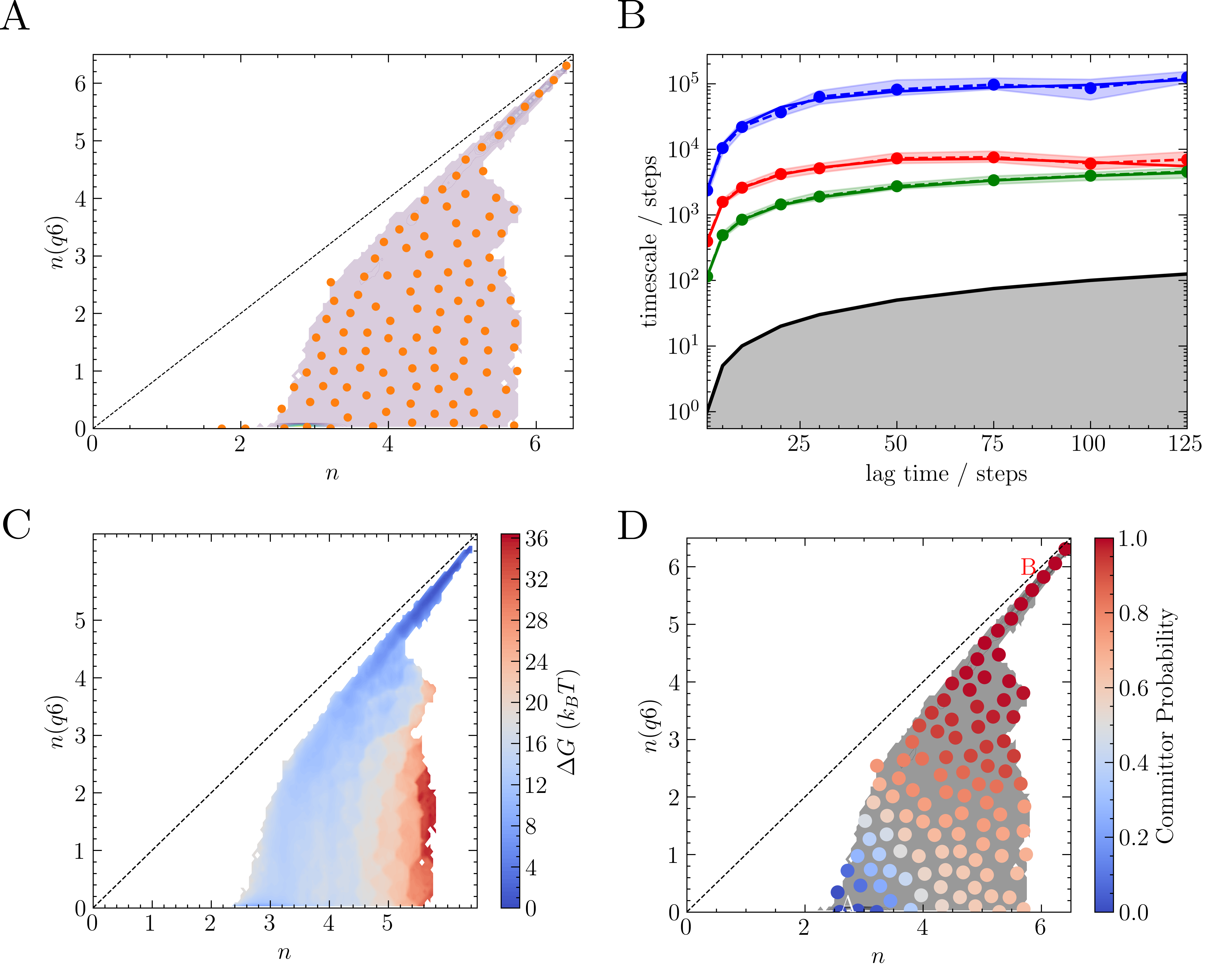}
    \caption{State points (A), lag times (B), free energies (C) and committor probabilities (D) from the MSM characterising transitions between states in $n,n(q6)$.}
    \label{fig:msm-box}
\end{figure*}

\clearpage

\end{document}